\documentclass[12pt]{article}
\usepackage{graphicx}

\def\hybrid{\topmargin 0pt      \oddsidemargin 0pt
        \headheight 0pt \headsep 0pt
       \voffset-1cm
        \textwidth 6.25in       
       \textheight 9.5in       
        \marginparwidth 0.0in
        \parskip 5pt plus 1pt   \jot = 1.5ex}
\catcode`\@=11
\def\marginnote#1{}

\newcount\hour
\newcount\minute
\newtoks\amorpm
\hour=\time\divide\hour by60
\minute=\time{\multiply\hour by60 \global\advance\minute by-\hour}
\edef\standardtime{{\ifnum\hour<12 \global\amorpm={am}%
        \else\global\amorpm={pm}\advance\hour by-12 \fi
        \ifnum\hour=0 \hour=12 \fi
        \number\hour:\ifnum\minute<10 0\fi\number\minute\the\amorpm}}
\edef\militarytime{\number\hour:\ifnum\minute<10 0\fi\number\minute}

\def\draftlabel#1{{\@bsphack\if@filesw {\let\thepage\relax
   \xdef\@gtempa{\write\@auxout{\string
      \newlabel{#1}{{\@currentlabel}{\thepage}}}}}\@gtempa
   \if@nobreak \ifvmode\nobreak\fi\fi\fi\@esphack}
        \gdef\@eqnlabel{#1}}
\def\@eqnlabel{}
\def\@vacuum{}
\def\draftmarginnote#1{\marginpar{\raggedright\scriptsize\tt#1}}

\def\draftlabel#1{{\@bsphack\if@filesw {\let\thepage\relax
   \xdef\@gtempa{\write\@auxout{\string
      \newlabel{#1}{{\@currentlabel}{\thepage}}}}}\@gtempa
   \if@nobreak \ifvmode\nobreak\fi\fi\fi\@esphack}
        \gdef\@eqnlabel{#1}}
\def\@eqnlabel{}
\def\@vacuum{}
\def\draftmarginnote#1{\marginpar{\raggedright\scriptsize\tt#1}}

\def\draft{\oddsidemargin -.5truein
        \def\@oddfoot{\sl preliminary draft \hfil
        \rm\thepage\hfil\sl\today\quad\militarytime}
        \let\@evenfoot\@oddfoot \overfullrule 3pt
        \let\label=\draftlabel
        \let\marginnote=\draftmarginnote
   \def\@eqnnum{(\theequation)\rlap{\kern\marginparsep\tt\@eqnlabel}%
\global\let\@eqnlabel\@vacuum}  }

\def\numberbysection{\@addtoreset{equation}{section}
        \def\theequation{\thesection.\arabic{equation}}}

\def\underline#1{\relax\ifmmode\@@underline#1\else
        $\@@underline{\hbox{#1}}$\relax\fi}

\def\titlepage{\@restonecolfalse\if@twocolumn\@restonecoltrue\onecolumn
     \else \newpage \fi \thispagestyle{empty}\c@page\z@
        \def\thefootnote{\fnsymbol{footnote}} }

\def\endtitlepage{\if@restonecol\twocolumn \else  \fi
        \def\thefootnote{\arabic{footnote}}
        \setcounter{footnote}{0}}  
\relax

\numberbysection
\hybrid

\newfont{\Bbb}{msbm10 scaled 1\@ptsize00}
\newfont{\Bbbb}{msbm7 scaled 1\@ptsize00}

\newcommand{\DDD}{\raise-1pt\hbox{$\mbox{\Bbbb D}$}}

\newcommand{\UUU}{\raise-1pt\hbox{$\mbox{\Bbbb U}$}}

\newcommand{\ZZ}{\mbox{\Bbb Z}}
\newcommand{\z}{\raise-1pt\hbox{$\mbox{\Bbbb Z}$}}

\newcommand{\sss}{\raise-1pt\hbox{$\mbox{\Bbbb S}$}}

\def\beq{\begin{equation}}
\def\eeq{\end{equation}}

\newtheorem{theorem}{Theorem}[section]

\newtheorem{lemma-definition}{Lemma-Definition}[section]

\begin{document}

\begin{titlepage}

\title{Elliptic Cauchy matrices}

\author{V. Prokofev\thanks{
Skolkovo Institute of Science and Technology, 143026, Moscow, Russia,
e-mail: vadprokofev@gmail.com}
\and
A.~Zabrodin\thanks{
Skolkovo Institute of Science and Technology, 143026, Moscow, Russia and
National Research University Higher School of Economics,
20 Myasnitskaya Ulitsa,
Moscow 101000, Russia and
NRC ``Kurchatov institute'', Moscow, Russia;
e-mail: zabrodin@itep.ru}}

\date{May 2023}
\maketitle

\vspace{-7cm} \centerline{ \hfill ITEP-TH-12/23}\vspace{7cm}

\begin{abstract}

Some identities that involve the elliptic version of the 
Cauchy matrices are presented and proved. They include the 
determinant formula, the formula for the inverse matrix, the
matrix product identity and the factorization formula.

\end{abstract}

\end{titlepage}

\vspace{5mm}

%

\tableofcontents

\vspace{5mm}

\section{Introduction}

Let ${\sf x}=\{x_1, \ldots , x_N\}$, 
${\sf y}=\{y_1, \ldots , y_N\}$ be two sets 
of pairwise distinct variables such that $x_i\neq y_j$ for all $i,j$.
The Cauchy matrix $C=C({\sf x}, {\sf y})$ is the $N\! \times \! N$
matrix with matrix elements
\beq\label{int1}
C_{ij}=\frac{1}{x_i-y_j}.
\eeq
The Cauchy matrix often appears in different applications (for example,
in soliton theory and theory of integrable systems).

The following result for determinant of $C$ belongs to
Cauchy and is well known:
\beq\label{int2}
\det_{1\leq i,j \leq N}C_{ij} =
\frac{\prod\limits_{a<b}(x_a-x_b)(y_b-y_a)}{\prod\limits_{a,b}
(x_a-y_b)}.
\eeq
The proof consists in considering both sides as rational functions 
of, say $x_1$, and identifying the residues at the simple poles $x_1=y_k$.
If all $x_i$'s and $y_i$'s are distinct, 
the determinant is 
nonzero and thus the Cauchy matrix is invertible.
The explicit formula for the inverse matrix is
\cite{Schechter}:
\beq\label{int3}
(C^{-1})_{ij}=\frac{(x_i-y_i)(x_j-y_j)}{x_j-y_i}
\, \prod_{k\neq i}\frac{y_i-x_k}{y_i-y_k}
\prod_{l\neq j}\frac{x_j-y_l}{x_j-x_l}.
\eeq

The elliptic version of the Cauchy matrix can be written using
the Weierstrass $\sigma$-function. 
Let $\omega$, $\omega '$ be complex numbers such that 
${\rm Im} (\omega '/ \omega )>0$.
The Weierstrass $\sigma$-function 
with quasi-periods $2\omega$, $2\omega '$ 
is defined by the following infinite product over the lattice
$2\omega m+2\omega ' m'$, $m,m'\in \ZZ$:
\beq\label{A1}
\sigma (x)=\sigma (x |\, \omega , \omega ')=
x\prod_{s\neq 0}\Bigl (1-\frac{x}{s}\Bigr )\, 
e^{\frac{x}{s}+\frac{x^2}{2s^2}},
\quad s=2\omega m+2\omega ' m' \quad m, m'\in \ZZ .
\eeq 
It is an odd entire quasiperiodic function in the complex plane. 
As $x\to 0$,
$\sigma (x)=x+O(x^5)$.
The monodromy properties of the $\sigma$-function 
under shifts by the quasi-periods
are as follows:
\beq\label{A4}
\begin{array}{l}
\sigma (x+2\omega )=-e^{2\eta (x+\omega )}\sigma (x),
\\ \\
\sigma (x+2\omega ' )=-e^{2\eta '(x+\omega ' )}\sigma (x).
\end{array}
\eeq
Here $\eta , \eta '$ are constants which can be expressed 
in terms of the Weierstrass $\zeta$-function $\zeta (x)$ defined as
$\zeta (x)=\sigma '(x)/\sigma (x)$; we have: $\eta =\zeta (\omega )$, 
$\eta '=\zeta (\omega ')$. The constants $\eta , \eta '$ are connected
by the important relation
\beq\label{}
2\eta \omega ' -2\eta '\omega =\pi i.
\eeq
The monodromy properties imply that the function
$$
f(x)=\prod_{\alpha =1}^M \frac{\sigma 
(x-a_{\alpha})}{\sigma (x-b_{\alpha})}, \qquad \sum_{\alpha =1}^M
(a_{\alpha}-b_{\alpha})=0
$$
is a double-periodic function with periods $2\omega$, $2\omega '$
(an elliptic function). If $\omega , \omega '$ tend to infinity, then
$\sigma (x)\to x$. For the Weierstrass and other elliptic functions
see \cite{WW,Akhiezer}.

The elliptic generalization of the Cauchy matrix depends on an 
additional parameter $\lambda$:
\beq\label{int4}
C_{ij}({\sf x}, {\sf y}; \lambda )=
\frac{\sigma (x_i-y_j+\lambda )}{\sigma (\lambda )\sigma (x_i-y_j)}.
\eeq
The rational Cauchy matrix (\ref{int1}) is obtained from it in the 
double limit $\omega , \omega '\to \infty$ and then $\lambda \to \infty$.
The elliptic Cauchy matrices play an important role in the theory 
of the elliptic Ruijsenars-Schneider systems \cite{RS86,Ruij87}
(relativistic analogues of the Calogero-Moser systems).

The aim of this paper is to present some identities for the elliptic 
Cauchy matrix which generalize (\ref{int2}), (\ref{int3}).
For completeness, we also present the factorization formula and
Gauss decomposition of the Cauchy matrices. 

\section{The main identities}

\paragraph{The determinant.}
The expression for the determinant of $C({\sf x}, {\sf y}; \lambda )$
is due to Frobenius \cite{Frobenius}:
\beq\label{id1}
\det_{1\leq i,j \leq N}
C_{ij}({\sf x}, {\sf y}; \lambda )=
\frac{\sigma (\lambda +X-Y)}{\sigma (\lambda )}\, 
\frac{\prod\limits_{a<b}\sigma (x_a-x_b)\sigma 
(y_b-y_a)}{\prod\limits_{a,b}
\sigma (x_a-y_b)},
\eeq
where
$$
X=\sum_{i=1}^Nx_i, \qquad Y=\sum_{i=1}^Ny_i.
$$

\paragraph{The inverse matrix.}
The inverse matrix is
\beq\label{id2}
\begin{array}{l}
\displaystyle{
(C^{-1}({\sf x}, {\sf y}; \lambda ))_{ij}=
\frac{\sigma (y_i\! -\! x_j\! +\! 
\lambda \! +\! X\! -\! Y)}{\sigma (\lambda )}\,
\frac{\sigma (x_i-y_i)\sigma (x_j-y_j)}{\sigma (x_j-y_i)}}
\\ \\
\phantom{aaaaaaaaaaaaaaaaaaaaaa}\displaystyle{
\times \prod_{k\neq i}\frac{\sigma (y_i-x_k)}{\sigma (y_i-y_k)}
\prod_{l\neq j}\frac{\sigma (x_j-y_l)}{\sigma (x_j-x_l)}.}
\end{array}
\eeq

\paragraph{The matrix product identity.}
Next we present an identity which generalizes (\ref{id2}).
Let $D({\sf x}, {\sf y})$ be the diagonal matrix with matrix elements
\beq\label{id3}
D_{ij}({\sf x}, {\sf y})=\delta_{ij}\sigma (x_i-y_i)
\prod_{k\neq i}\frac{\sigma (x_i-y_k)}{\sigma (x_i-x_k)}.
\eeq
Introduce the matrix
\beq\label{id4}
G_{\lambda}({\sf x}, {\sf y})=D({\sf x}, {\sf y})
C({\sf x}, {\sf y}; \lambda ).
\eeq
In particular, $G_{\lambda}({\sf x}, {\sf x})=I$, where $I$ 
is the unity matrix\footnote{The matrix 
$G_{\lambda}({\sf x}, {\sf x})$ should be understood as
$\lim\limits_{\varepsilon \to 0}G_{\lambda}({\sf x}, 
{\sf x}_{\varepsilon})$,
where ${\sf x}_{\varepsilon}=\{x_1+\varepsilon , 
\ldots , x_N+\varepsilon \}$.}.

\begin{theorem}\label{theorem:product}
The following identity holds true:
\beq\label{id5}
G_{\lambda +Y}({\sf x}, {\sf y})G_{\lambda +Z}({\sf y}, {\sf z})=
G_{\lambda +Z}({\sf x}, {\sf z}),
\eeq
where ${\sf z}=\{ z_1, \ldots , z_N\}$ is a third set of variables
and $\displaystyle{Z=\sum_{i=1}^N z_i}$.
\end{theorem}

\noindent
At ${\sf z}={\sf x}$ we get
$G_{\lambda +Y}({\sf x}, {\sf y})G_{\lambda +X}({\sf y}, {\sf x})=I$
which is (\ref{id2}). We call (\ref{id5}) the matrix product identity. 
It was used for the analysis of elliptic solutions to the Toda 
hierarchy with constraint of type B
in our recent paper \cite{PZ23}.

Note that
$$
C^t ({\sf x}, {\sf y}; \lambda )=-C ({\sf y}, {\sf x}; -\lambda ),
$$
where $C^t$ is the transposed matrix.
The transposition of identity (\ref{id5}) gives the identity
\beq\label{id6}
H_{\lambda -X}({\sf x}, {\sf y})H_{\lambda -Y}({\sf y}, {\sf z})=-
H_{\lambda -X}({\sf x}, {\sf z})
\eeq
for the matrix
\beq\label{id7}
H_{\lambda}({\sf x}, {\sf y})=C({\sf x}, {\sf y}; \lambda )
D({\sf y}, {\sf x}).
\eeq

In the rational limit $\omega , \omega '\to \infty$, $\lambda \to \infty$
we have
\beq\label{id8a}
G_{\lambda }({\sf x}, {\sf y})\to K({\sf x}, {\sf y}), \qquad
K_{ij}({\sf x}, {\sf y})=
\frac{\prod\limits_{k\neq j}(x_i-y_k)}{\prod\limits_{l\neq i}(x_i-x_l)}
\eeq
and the matrix product identity becomes the matrix equality
\beq\label{id8}
K({\sf x}, {\sf y})K({\sf y}, {\sf z})=K({\sf x}, {\sf z}).
\eeq

\paragraph{The factorization formula.} The matrix product identity
suggests that there exists a matrix $g_{\lambda}({\sf x})$ such that
$G_{\lambda +Y}({\sf x}, {\sf y})=g_{\lambda}({\sf x})
g^{-1}_{\lambda}({\sf y})$.
This is indeed the case.

\begin{theorem}\label{theorem:factorization}
The matrix $G_{\lambda}({\sf x}, {\sf y})$ admits a factorization
of the form
\beq\label{f1}
G_{\lambda +Y}({\sf x}, {\sf y})=g_{\lambda}({\sf x})
g^{-1}_{\lambda}({\sf y}).
\eeq
Here $g_{\lambda}({\sf x})$ is the matrix
\beq\label{f2}
(g_{\lambda}({\sf x}))_{ij}=
\frac{\sigma^{(j)}(x_i+\frac{\lambda}{N})}{\prod\limits_{l\neq i}
\sigma (x_i-x_l)},
\eeq
where
\beq\label{f3}
\sigma^{(k)}(x)=e^{2\eta 'kx}\prod_{l=0}^{N-1}
\sigma \Bigl (x+\frac{N\! -\! 2l\! -\! 1}{N}\, \omega -\frac{2k}{N}\,
\omega ' \Bigr ).
\eeq
\end{theorem}

\noindent
Clearly, there is a freedom in the definition of the 
matrix $g_{\lambda}$: it can be multiplied from the right by arbitrary
non-degenerate constant matrix $S$: $g_{\lambda}\to g_{\lambda}S$ .
We call (\ref{f1}) the factorization formula. 

Note that factorization
of the closely related 
Lax matrix for the Ruijsenaars-Schneider model (which is 
the Cauchy matrix with $y_i=x_i+c$ multiplied by a diagonal matrix) 
was discussed in \cite{VZ}. It should be also mentioned 
that the factorization formula has a 
``quantum analogue'' as factorization of the quantum $L$-operator 
of generalized quantum spin chains \cite{H97}. 

Let us present trigonometric and rational degenerations of the 
factorization formula. These degenerations are non-trivial because 
a naive limit of the matrix $g_{\lambda}({\sf x})$ leads to 
a degenerate matrix for which the inverse matrix does not exist. 
Nevertheless, the factorization formulas do exist in these cases. 

In the trigonometric limit $\omega '\to i\infty$. Put $\omega =\pi /2$,
then $\sigma (x|\frac{\pi}{2}, i\infty )=e^{x^2/6}\sin x$. It is easy to
see that the matrix product identity (\ref{id5}) holds for the 
trigonometric version of the matrix $G_{\lambda}({\sf x}, {\sf y})$:
\beq\label{trig1}
(G_{\lambda}^{\rm trig}({\sf x}, {\sf y}))_{ij}=
\sin (x_i-y_i)\left (
\prod_{l\neq i}\frac{\sin (x_i-y_l)}{\sin (x_i-x_l)}\right )
\frac{\sin (x_i-y_j+\lambda )}{\sin \lambda \, \sin (x_i-y_j)}.
\eeq

\begin{theorem}\label{theorem:factorization-trig}
In the trigonometric case the factorization formula holds in the form
\beq\label{trig2}
G_{\lambda}^{\rm trig}({\sf x}, {\sf y})=
g_{\lambda}^{\rm trig}({\sf x})
(g^{\rm trig}_{\lambda}({\sf y}))^{-1},
\eeq
where
\beq\label{trig3}
(g_{\lambda}^{\rm trig}({\sf x}))_{jk}=
\frac{\varphi_k(x_j)}{\prod\limits_{l\neq j} \sin (x_j-x_l)}
\eeq
and
\beq\label{trig4}
\varphi_k(x_j)=e^{-iNx_j}\Bigl (e^{2ik(x_j+\frac{\lambda}{N})}+
(-1)^N \delta_{kN}\Bigr ), \quad j,k=1, \ldots , N.
\eeq
\end{theorem}

In the rational limit $\omega \to \infty,\, \omega '\to i\infty$ and
$\sigma (x|\infty , i\infty )=x$. The matrix 
\beq\label{rat1}
(G_{\lambda}^{\rm rat}({\sf x}, {\sf y}))_{ij}=
(x_i-y_i)\left (
\prod_{l\neq i}\frac{x_i-y_l}{x_i-x_l}\right )
\frac{x_i-y_j+\lambda }{\lambda \, (x_i-y_j)}
\eeq
is obtained from (\ref{trig1}) after the substitutions 
$x_i\to \epsilon x_i$, 
$y_i\to \epsilon y_i$, $\lambda \to \epsilon \lambda$ in the limit
$\epsilon \to 0$. 

\begin{theorem}\label{theorem:factorization-rat}
In the rational case the factorization formula holds in the form
\beq\label{rat2}
G_{\lambda}^{\rm rat}({\sf x}, {\sf y})=
g_{\lambda}^{\rm rat}({\sf x})
(g^{\rm rat}_{\lambda}({\sf y}))^{-1},
\eeq
where
\beq\label{rat3}
(g_{\lambda}^{\rm rat}({\sf x}))_{jk}=\frac{(x_{j} +
\frac{\lambda}{N})^{k-1 +\delta_{kN}}}{\prod\limits_{l\neq j}
(x_j-x_l)}, \quad
j,k=1, \ldots , N.
\eeq
\end{theorem}

In the limit $\lambda \to \infty$ the factorization formula 
for the matrix $K({\sf x}, {\sf y})$ given by (\ref{id8a}) 
holds in the following form:
\beq\label{rat4}
K({\sf x}, {\sf y})=W({\sf x})W^{-1}({\sf y}),
\eeq
where
\beq\label{rat5}
W_{ij}({\sf x})=\frac{x_i^{j-1}}{\prod\limits_{l\neq i}
(x_i-x_l)}, \quad i,j=1, \ldots , N.
\eeq

\paragraph{The Gauss decomposition.}
The Gauss decomposition of the elliptic Cauchy matrix was found in
\cite{FKR11}. We present it here in a different form. 

\begin{theorem}\hspace{-2mm}(\cite{FKR11})\label{theorem:gauss}
The Gauss
decomposition of the elliptic Cauchy matrix 
$C=C({\sf x}, \! {\sf y}, \! \lambda )$ has the form
\beq\label{g1}
C=UDL,
\eeq
where $U$ and $L$ are respectively upper and lower triangular matrices
with $1$ on the main diagonal and $D$ is a diagonal matrix. These matrices
are as follows:
\beq\label{g2}
\begin{array}{l}
\displaystyle{
U_{ij}=\frac{\sigma (x_i-y_j+\lambda_j)
\sigma (x_j-y_j)}{\sigma (x_j-y_j+\lambda_j)\sigma (x_i-y_j)}
\prod_{l=j+1}^N\frac{\sigma (x_i-x_l)\sigma (x_j-y_l)}{\sigma (x_j-x_l)
\sigma (x_i-y_l)}, \quad i\leq j,}
\\ \\
\displaystyle{
D_{jj}=\frac{\sigma (x_j-y_j+
\lambda_j)}{\sigma (\lambda_j)\sigma (x_j-y_j)}
\prod_{l=j+1}^N\frac{\sigma (x_j-x_l)\sigma (y_j-y_l)}{\sigma (x_j-y_l)
\sigma (y_j-x_l)},}
\\ \\
\displaystyle{
L_{jk}=\frac{\sigma (x_j-y_k+\lambda_j)
\sigma (x_j-y_j)}{\sigma (x_j-y_j+\lambda_j)\sigma (x_j-y_k)}
\prod_{l=j+1}^N\frac{\sigma (y_k-y_l)\sigma (y_j-x_l)}{\sigma (y_k-x_l)
\sigma (y_j-y_l)}, \quad j\geq k,}
\end{array}
\eeq
where
\beq\label{g3}
\lambda_j=\lambda +\sum_{l=j+1}^N(x_l-y_l), \quad \lambda_N \equiv \lambda .
\eeq
\end{theorem}

\noindent
In (\ref{g2}), the products in the case $j=N$ should be put equal to $1$. 

\section{Proofs}

\paragraph{Proof of the determinant formula (\ref{id1}).}
The proof of (\ref{id1}) is standard. 
Consider the ratio $r$ of the left
and right hand sides as a function of $x_1$. It is easy to see that 
it is an elliptic function of $x_1$. Possible (simple) poles may occur
at $x_1=x_k$, $k=2, \ldots , N$. However, these poles are absent
because the numerator (the determinant) also vanishes 
at these points. Another pole
can occur 
when $\lambda +X-Y=0$ but it is impossible for 
elliptic function to have a single
simple pole in the fundamental domain 
and thus there is no pole at this point. So the function $r$
is regular and thus
does not depend on $x_1$. In the same way one can prove that it does
not depend on all $x_i$'s and $y_i$'s, and thus it is a constant.
The constant can be found setting $x_i\to \varepsilon x_i$, 
$y_i\to \varepsilon y_i$ and tending $\varepsilon \to 0$. Then
$\sigma (\varepsilon x)$ can be substituted by $\varepsilon x$ 
in the leading order and,
using (\ref{int2}), one can see that the constant is equal to $1$.
 
\paragraph{Proof of the formula for inverse matrix (\ref{id2}).}
Formula (\ref{id2}) is a consequence of (\ref{id1}) since any minor
of the elliptic Cauchy matrix is again determinant of an elliptic 
Cauchy matrix and can be found using (\ref{id1}). 

\paragraph{Proof of Theorem \ref{theorem:product}.}
Now let us prove (\ref{id5}). We will prove it in the form
$$
G_{\lambda}({\sf x}, {\sf y})G_{\lambda -\delta}({\sf y}, {\sf z})=
G_{\lambda -\delta}({\sf x}, {\sf z}),
$$
where $\delta = Y-Z$. Consider the function
$$
S_{ik}:=\frac{(G_{\lambda}({\sf x}, {\sf y})
G_{\lambda -\delta}({\sf y}, {\sf z})_{ik}}{(G_{\lambda -
\delta}({\sf x}, {\sf z}))_{ik}}.
$$
Explicitly, we have:
\beq\label{p1}
\begin{array}{c}
\displaystyle{
S_{ik}=\frac{\sigma (x_i-y_i)}{\sigma (x_i-z_i)}
\prod_{l\neq i} \frac{\sigma (x_i-y_l)}{\sigma (x_i-z_l)}}
\\ \\
\displaystyle{\times 
\sum_j \frac{\sigma (x_i \! - \! z_k)\sigma (y_j \! - \! z_j)
\sigma (x_i \! - \! y_j \! +\! \lambda )
\sigma (y_j \! - \! z_k \! -\! \delta \! +\! \lambda )}{\sigma (\lambda )
\sigma (x_i \! - \! z_k \! -\! \delta \! +\! \lambda )
\sigma (x_i \! - \! y_j )\sigma (y_j \! - \! z_k)}
\prod_{s\neq j} \frac{\sigma (y_j-z_s)}{\sigma (y_j-y_s)}}.
\end{array}
\eeq
This function is obviously elliptic in $\lambda$. It has two possible
simple poles at $\lambda =0$ and $\lambda =z_k-x_i+\delta$ (if
$z_k-x_i+\delta \neq 0$). The residues are proportional to
$$
R=\sum_j \frac{\sigma (y_j-z_j)\sigma (y_j-z_k-\delta )}{\sigma (y_j-z_k)}
\prod_{s\neq j} \frac{\sigma (y_j-z_s)}{\sigma (y_j-y_s)}.
$$
But this is nothing else than the sum of residues of the elliptic
function
$$
f(y)=\frac{\sigma (y-z_k-\delta )}{\sigma (y-z_k)}
\prod_{s} \frac{\sigma (y-z_s)}{\sigma (y-y_s)}
$$
and thus is equal to $0$. When $z_k-x_i+\delta =0$ we have a double
pole with coefficient proportional to $R=0$. So $S_{ik}$ is
a regular elliptic function of $\lambda$ and thus it does not depend 
on $\lambda$. 

Consider now $S_{ik}$ as a function of $x_i$. It is easy to see that 
it is an elliptic function of $x_i$. A possible pole may lie at
$x_i=z_k -\delta +\lambda$ but actually this pole is absent because
it depends on $\lambda$. The other poles may lie at $x_i=z_l$ with
$l\neq k$; the residue at such a pole is proportional to
$$
J_l=\sum_j \frac{\sigma (z_l-y_j+\lambda )
\sigma (y_j-z_k-\delta +\lambda )\sigma (y_j-z_j)}{\sigma (z_l-y_j)
\sigma (y_j-z_k)}\prod_{s\neq j}
\frac{\sigma (y_j-z_s)}{\sigma (y_j-y_s)}
$$
$$
=(C({\sf z}, {\sf y}; \lambda )
G_{\lambda -\delta}({\sf y}, {\sf z}))_{lk}
=(D^{-1}({\sf z}, {\sf y}))_{lk}=0
$$
because $D({\sf z}, {\sf y})$ is a diagonal matrix and $l\neq k$. 
So $S_{ik}$ is regular in $x_i$ and thus does not depend on $x_i$.
Its value can be found by putting $x_i=y_i$, then in the sum 
in the right hand side of (\ref{p1}) only the term at $j=i$ 
survives and
after cancellations we get $S_{ik}=1$. 

Let us give another proof of the matrix product
identity (\ref{id5}). Consider the 
linear space of functions $\psi (x)$ having simple poles at the 
$N$ points $x_1, \ldots , x_N$ in the fundamental domain and such that
$\psi (x+2\omega )=e^{2\eta \lambda}\psi (x)$, 
$\psi (x+2\omega ')=e^{2\eta '\lambda}\psi (x)$ with some 
$\lambda$.
Such functions are called double-Bloch functions. It is easy to see 
that this linear space is $N$-dimensional. Functions from this space
can be written as
\beq\label{p2}
\psi (x)=\sum_{i=1}^N c_i \frac{\sigma (x-x_i+\lambda )}{\sigma (\lambda )
\sigma (x-x_i)},
\eeq
where $c_i$ are arbitrary coefficients. Alternatively, such a function
can be parametrized through its zeros $u_i$:
\beq\label{p3}
\psi (x)=C \prod_{i=1}^N \frac{\sigma (x-u_i)}{\sigma (x-x_i)},
\eeq
where $u_i$ are subject to the condition
\beq\label{p4}
\sum_i u_i :=U =X-\lambda .
\eeq
Consider now the function
$$
\psi_1 (x)=\psi (x)\prod_i \frac{\sigma (x-x_i)}{\sigma (x-y_i)}=
\prod_i \frac{\sigma (x-u_i)}{\sigma (x-y_i)}.
$$
It can be expanded as
$$
\psi_1(x)=\sum_i b_i \frac{\sigma 
(x-y_i +\lambda -X+Y)}{\sigma (\lambda -X+Y)\sigma 
(x-y_i)},
$$
and the vectors ${\bf c}=(c_1, \ldots , c_N)^t$ and 
${\bf b}=(b_1, \ldots , b_N)^t$ are connected as
$$
{\bf b}=G_{\lambda}({\sf y}, {\sf x}){\bf c}.
$$
Next, consider the function
$$
\psi_2 (x)=\psi_1 (x)\prod_i \frac{\sigma (x-y_i)}{\sigma (x-z_i)}=
\prod_i \frac{\sigma (x-u_i)}{\sigma (x-z_i)}.
$$
It can be expanded as
$$
\psi_2(x)=\sum_i a_i \frac{\sigma 
(x-z_i +\lambda -X+Z)}{\sigma (\lambda -X+Z)\sigma 
(x-z_i)}
$$
with some coefficients $a_i$, with 
${\bf a}=G_{\lambda}({\sf z}, {\sf x}){\bf c}$. On the other hand,
$$
{\bf a}=G_{\lambda -X+Y}({\sf z}, {\sf y}){\bf b}=
G_{\lambda -X+Y}({\sf z}, {\sf y})G_{\lambda}({\sf y}, {\sf x}){\bf c}.
$$
Since the vector ${\bf c}$ is arbitrary, we have
$$
G_{\lambda -X+Y}({\sf z}, {\sf y})G_{\lambda}({\sf y}, {\sf x})
=G_{\lambda}({\sf z}, {\sf x})
$$
which is equivalent to (\ref{id5}).

Finally, let us note that the matrix product identity (\ref{id5})
immediately follows from the factorization formula to be
proved below. 

\paragraph{Proof of the elliptic factorization formula.}
We will prove (\ref{f1}) in the form
\beq\label{pf1}
G_{\lambda +Y}({\sf x}, {\sf y})g_{\lambda}({\sf y})=g_{\lambda}({\sf x}),
\eeq
or
\beq\label{pf2}
\frac{\sigma (x_i-y_i)}{\sigma (\lambda +Y)}
\prod_{l\neq i} \frac{\sigma (x_i-y_l)}{\sigma (x_i-x_l)}
\sum_j \frac{\sigma (\lambda \! +\! Y \! +\! x_i \! -\! 
y_j)}{\sigma (x_i-y_j)}
\frac{\sigma^{(k)}(y_j+\frac{\lambda}{N})}{\prod\limits_{l\neq j}
\sigma (y_j-y_l)}
=\frac{\sigma^{(k)}(x_i+\frac{\lambda}{N})}{\prod\limits_{l\neq i}
\sigma (x_i-x_l)},
\eeq
where we have substituted the explicit form of the matrices
$g_{\lambda}({\sf x})$, $g_{\lambda}({\sf y})$ (\ref{f2}). 
Consider the left hand side of (\ref{pf2}) as a function of $y_i$
and denote it as $f(y_i)$. 
The terms of the sum with $j\neq i$ are obviously elliptic functions
of $y_i$. As far as the term with $j=i$ is concerned, we need the 
monodromy properties of the function $\sigma^{(k)}(x)$ which can be
obtained with the help of (\ref{f3}):
\beq\label{pf3}
\begin{array}{l}
\sigma^{(k)}(x+2\omega )=(-1)^N e^{2\eta N(x+\omega )}
\sigma^{(k)}(x),
\\ \\
\sigma^{(k)}(x+2\omega ')=(-1)^N e^{2\eta 'N(x+\omega ')}
\sigma^{(k)}(x).
\end{array}
\eeq
As it is seen from (\ref{pf2}), the function $f(y_i)$ may have simple 
poles
at $y_i=y_l$ ($l\neq i$) and at the point where $\lambda +Y=0$.  
It is easily verified that residues at the former $N-1$ poles vanish.
Since any non-constant elliptic function must have at least two 
simple poles, the pole at $\lambda +Y=0$ is actually absent, and the 
function $f(y_i)$ turns out to be regular. Therefore, it does not
depend on $y_i$ and its value can be found by putting $y_i$ equal to
some particular value. It is convenient to put $y_i=x_i$, then only the
term with $j=i$ in the sum survives, and the result coincides with the
right hand side. 

\paragraph{Proof of the factorization formula in the trigonometric case.}
We will prove the factorization formula (\ref{trig2}) in the form
$$
G_{\lambda}^{\rm trig}({\sf x}, {\sf y})
g^{\rm trig}_{\lambda}({\sf y})=
g_{\lambda}^{\rm trig}({\sf x}).
$$
With $g_{\lambda}^{\rm rat}$ as in (\ref{trig3}), this is equivalent
to the equality
\beq\label{ptf1}
\sum_{j=1}^N \left (\prod_{l\neq j}^N \frac{\sin (x-y_l)}{\sin (y_j-y_l)}
\right ) \frac{\sin (x-y_j +\lambda +Y)}{\sin (\lambda +Y)}\,
\varphi_k (y_j)=\varphi_k(x),
\eeq
where $\varphi_k$ is given by (\ref{trig4}). It is convenient to change
the variables as
$$
z=e^{2ix}, \quad w_j=e^{2iy_j}, \quad t=e^{2i(\lambda +Y)},
$$
then (\ref{ptf1}) acquires the form
\beq\label{ptf2}
\sum_{j=1}^N \left (\prod_{l\neq j}^N \frac{z-w_l}{w_j-w_l}\right )
\frac{zw_j^{-1}t -1}{t-1}\, \tilde \varphi_k (w_j)=\tilde \varphi_k (z),
\eeq
where
\beq\label{ptf3}
\tilde \varphi_k (z)=z^ke^{2ik\lambda /N}+(-1)^N\delta_{kN}, \quad
k=1, \ldots , N.
\eeq
Consider first the case $k=1, \ldots , N-1$. Then both sides of
(\ref{ptf2}) are polynomials in $z$ of degree strictly less than $N$.
Indeed, since
$$
\sum_{j=1}^N \frac{w_j^k}{\prod\limits_{l\neq j}(w_j-w_l)}=0
\quad \mbox{if $\; k=0, \ldots , N-2$},
$$
the coefficient in front of the highest term $z^N$ in the left hand side
is equal to $0$. The values of both sides at the $N$ points
$w_1, \ldots , w_N$ are the same. Therefore, the polynomials coincide. 
At $k=N$ we have $\tilde \varphi_N (z)=z^N e^{2i\lambda}+(-1)^N$.
The both sides of (\ref{ptf2}) are polynomials in $z$ of degree
$N$ and their values at the $N$ points
$w_1, \ldots , w_N$ are the same. In order to prove that the polynomials
coincide, it is enough to compare the coefficients in front of the
highest terms $z^N$ are the same. The highest coefficient in the left hand
side is equal to
$$
a_N=\frac{e^{2i\lambda}t}{t-1}\sum_{j=1}^N 
\frac{w_j^{N-1}}{\prod\limits_{l\neq j}(w_j-w_l)}+\frac{(-1)^N t}{t-1}
\sum_{j=1}^N 
\frac{w_j^{-1}}{\prod\limits_{l\neq j}(w_j-w_l)}.
$$
As is easy to see,
$$
\sum_{j=1}^N 
\frac{w_j^{N-1}}{\prod\limits_{l\neq j}(w_j-w_l)}=1 \quad 
\mbox{and} \quad
\sum_{j=1}^N 
\frac{w_j^{-1}}{\prod\limits_{l\neq j}(w_j-w_l)}=(-1)^{N-1}
(w_1\ldots w_N)^{-1}. 
$$
Therefore, we conclude that $a_N=e^{2i\lambda}$ is the same as 
in the right hand side.

\paragraph{Proof of the factorization formula in the rational case.}
We will prove the factorization formula (\ref{rat2}) in the form
$$
G_{\lambda}^{\rm rat}({\sf x}, {\sf y})
g^{\rm rat}_{\lambda}({\sf y})=
g_{\lambda}^{\rm rat}({\sf x}).
$$
With $g_{\lambda}^{\rm rat}$ as in (\ref{rat3}), this is equivalent
to the equality
\beq\label{prf1}
\sum_{j=1}^N \Bigl (1+\frac{x-y_j}{\lambda +Y}\Bigr )
\Bigl (y_j +\frac{\lambda}{N}\Bigr )^{k-1+\delta_{kN}}
\prod_{l\neq j}^N\frac{x-y_l}{y_j-y_l}=
\Bigl (x +\frac{\lambda}{N}\Bigr )^{k-1+\delta_{kN}}
\eeq
which is an identity. Indeed, the left hand side 
is a polynomial in $x$ of degree not greater than $N$. Let us denote it by
$P^{(k)}_N(x)$. The values of this polynomial at $N$ points $y_j$ are
$$
P^{(k)}_N(y_j)=\Bigl (y_j +\frac{\lambda}{N}\Bigr )^{k-1+\delta_{kN}}.
$$
As $x\to \infty$ we have 
$$
P^{(k)}_N(x)=c_k x^N + O(x^{N-1}), \quad
c_k=\frac{1}{\lambda +Y}\sum_j 
\frac{(y_j +\frac{\lambda}{N})^{k-1+
\delta_{kN}}}{\prod\limits_{l\neq j}(y_j-y_l)}.
$$
It is not difficult to see that $c_k=0$ at $k=1,\ldots , N-1$ and
$c_N=1$. Indeed,
$$
\sum_{j=1}^N
\frac{y_j^k}{\prod\limits_{l\neq j}^{N}(y_j-y_l)}=0 \quad 
\mbox{if $\; 0\leq k \leq N-2$}, \quad
\mbox{and} \quad
\sum_{j=1}^N
\frac{y_j^{N}}{\prod\limits_{l\neq j}^{N}(y_j-y_l)}=\sum_{j=1}^Ny_j.
$$
Therefore, at $k=1,\ldots , N-1$ 
the both sides of (\ref{prf1}) are polynomials of 
degree strictly less than $N$ whose values are the same at $N$ points;
thus they coincide. In the case $k=N$ the both sides are polynomials 
of degree $N$ with the same highest terms 
having the same values at $N$ points; thus they also coincide.

In the limit $\lambda \to \infty$ the matrix elements of 
$g_{\lambda}^{\rm rat}$ are singular. 
For the smooth limit, one should use
the freedom mentioned above and substitute the matrix 
$g_{\lambda}^{\rm rat}$ by $g_{\lambda}^{\rm rat}S(\lambda )$, where
$S(\lambda )$ is also singular as $\lambda \to \infty$ but the limit
of $g_{\lambda}^{\rm rat}S(\lambda )$ exists. In this way one obtains
the factorization of the matrix $K({\sf x}, {\sf y})$
of the form (\ref{rat4}), (\ref{rat5}). The factorization formula
is a direct consequence of the identity
$$
\sum_j y_j^k \prod_{l\neq j}\frac{x-y_l}{y_j-y_l}=x^k
$$
valid for $k=0,1, \ldots , N-1$.

\paragraph{Proof of Theorem \ref{theorem:gauss}.} The Gauss decomposition
was proved in \cite{FKR11}. Here we give another proof. 
Written in matrix elements, the Gauss decomposition (\ref{g1})
has the form
\beq\label{pg1}
C_{ik}=\sum_{j= {\rm max} (i,k)}^N U_{ij}D_{jj}L_{jk}.
\eeq
Substituting here the explicit formulas (\ref{g2}), we have:
\beq\label{pg2}
\begin{array}{l}
\displaystyle{
\frac{\sigma (x_i-y_k+\lambda )}{\sigma (\lambda )\sigma (x_i-y_k)}=
\! \sum_{j= {\rm max} (i,k)}^N \!\!
\frac{\sigma (x_j-y_j)\sigma (x_i\! -\! y_j\! +\! \lambda_j)
\sigma (x_j\! -\! y_k \! +\! \lambda_j)}{\sigma (\lambda_j)\sigma (\lambda_{j-1})
\sigma (x_i-y_j)\sigma (x_j-y_k)}}
\\ \\
\phantom{aaaaaaaaaaaaaaaaaaaaa}\displaystyle{ \times
\prod_{l=j+1}^N \frac{\sigma (x_i-x_l)\sigma (y_k-y_l)}{\sigma (x_i-y_l)
\sigma (y_k-y_l)}.}
\end{array}
\eeq
In order to prove this identity, let us consider both sides as
functions of $\lambda$. It is easy to see that the both sides are
double-Bloch functions of $\lambda$ with the Bloch multipliers 
$e^{2\eta (x_i-y_k)}$, $e^{2\eta '(x_i-y_k)}$. The function in the left hand
side has a single simple pole in the fundamental domain at $\lambda =0$
with residue $1$. Possible poles of the right hand side are at the
points where 
$$\lambda_j =\lambda +\sum_{l=j+1}^N (x_l-y_l)=0, \quad
\mbox{max}\, (i,k) \leq j \leq N.
$$
A direct verification shows that all these poles cancel except
the one at $\lambda_N=\lambda =0$, and the residue at this pole is equal
to $1$. Therefore, the functions in the left and right hand sides
coincide. 

\section*{Acknowledgments}

\addcontentsline{toc}{section}{Acknowledgments}

The work of A.Z. is an output of a research project 
implemented as a part of the Basic Research Program 
at the National Research University Higher School 
of Economics (HSE University).

\end{document}